\documentclass[a4paper,showpacs,twocolumn]{revtex4}
\usepackage{amssymb}
\usepackage{amsmath}
\usepackage{bm}
\usepackage{graphicx}
\usepackage{ulem}

\begin{document}
\title{CNOT gate by adiabatic passage with an optical cavity}
\date{\today}
\author{N. Sangouard$^{1}$}
\author{X. Lacour$^{2}$}
\author{S. Gu\'{e}rin$^{2}$}
\author{H. R. Jauslin$^{2}$}
\affiliation{$^{1}$ Fachbereich Physik, Universit\"at Kaiserslautern, Erwin-Schr\"odinger-Strasse D-67663 Kaiserlsautern, Germany.\\
$^{2}$ Laboratoire de Physique, Universit\'e de Bourgogne, UMR CNRS 5027, BP 47870,
21078 Dijon Cedex, France.}
\begin{abstract}
We propose a scheme for the construction of a CNOT gate by
adiabatic passage in an optical cavity. In opposition to a
previously proposed method, the technique is not based on
fractional adiabatic passage, which requires the control
of the ratio of two pulse amplitudes. Moreover, the technique
constitutes a decoherence-free method in the sense that
spontaneous emission and cavity damping are avoided since the
dynamics follows dark states.
\end{abstract}
\maketitle
\section{Introduction}
The controlled-not (CNOT) gate acts on systems composed of two
qubits. The first qubit controls the not operation on the second
(target) qubit : if the control qubit is in state $|0\rangle,$ the
target keeps its state whereas if the control is in state
$|1\rangle,$ the state of the target is switched. The set composed
of the CNOT gate and of elementary one-qubit gates forms a
universal set, i.e. all logical gates can be constructed
by the composition of gates in this set~\cite{Barenco}. The CNOT
gate allows to prepare entangled states from factorizable
superposition states. Entanglement is a key ingredient of quantum
computation \cite{galindo}, quantum teleportation~\cite{bennett}
or secure quantum cryptography~\cite{gisin} and thus confers to
the CNOT gate a broad practical interest. \\ The efficient
treatment of quantum information requires qubits insensitive to
decoherence, easily prepared and measured. Furthermore, the gates
operating on the qubits have to be robust with respect to
variations or partial knowledge of experimental parameters. These
requirements can be satisfied if the quantum information is
represented by atomic states controlled by adiabatic fields.
Indeed, the decoherence due to spontaneous emission can be avoided
if the dynamics follows dark states, i.e. states without
components on lossy excited states. In this context, a mechanism
has been proposed in Ref.~\cite{kis} to implement by adiabatic
passage all one-qubit gates, i.e. a general unitary matrix $U$ in
SU(2). A tripod-type system~\cite{unanyan} is used and as in
fractional stimulated Raman adiabatic passage
(f-STIRAP)~\cite{vitanov}, the amplitudes of two pulses
are required to have a constant ratio. The realisation of
this technique requires a specific system (for instance a system
of Zeeman states) to be robust~\cite{vitanov}. In a scheme, first
introduced in ref.~\cite{Pellizari}, composed of atoms fixed
inside a single-mode optical cavity, a mechanism has been proposed
\cite{Goto} for the creation of a two-qubit controlled-phase
(C-$phase$) gate by stimulated Raman adiabatic passage
(STIRAP) processes \cite{bergmann1, bergmann2} and a two-qubit
controlled-unitary (C-U) gate requiring f-STIRAP processes.
Additionally, it was suggested in ref.~\cite{Goto} to work with
five-level systems composed of three ground states (two of them
are the qubit states, the other one an ancillary state) and of two
excited states never populated in the adiabatic limit. The
proposal was to use the two excited states to realise the Raman
transitions involved respectively in the construction of one-qubit
gates and the C-$phase$ gate. This gives a technique for the
preparation of the universal set $\{U,$
C-$phase\}$~\cite{explanation} and all logical quantum gates can
thus be obtained from the composition of these two gates. The
construction of a CNOT gate from the universal set $\{U,$
C-$phase\}$ or from the C-U gate requires the control of the ratio
between two pulse amplitudes since f-STIRAP is used in both
methods. A five-level system in which the transitions can be
excited independently and the ratio
of the pulsed fields can be controlled robustly has to be found.\\
In this paper, we adapt to the preparation of the CNOT gate an
alternative mechanism based on adiabatic passage along dark states
that was used to construct directly the SWAP
gate~\cite{sangouard}. The mechanism is only based on
STIRAP processes. It can therefore be implemented robustly in a
variety of systems, avoiding e.g. the requirement encountered in
other schemes of using very specific Zeeman-sublevels. Moreover,
it constitutes a decoherence-free method in the sense that in the
adiabatic limit, the excited atomic states and the cavity mode (in
the limit of a cavity Rabi frequency much larger than the laser
Rabi frequency) are negligibly populated during the dynamics.
Furthermore, we also show that the proposed mechanism can be used
to directly prepare some specific composed gates. The usual
technique to construct a specific gate consists generally in
combining elementary gates belonging to a universal set. Since in
the experimental realisation of each gate there always are
uncontrollable losses, it is usefull to design instead direct
implementations
of specific compositions of elementary gates. \\
We present the atomic configuration associated to the qubits in
Section \ref{system}. In Section \ref{expo}, we develop the
mechanism and the analytical calculations of the instantaneous
eigenstates adiabatically involved in the dynamics. In Section
\ref{simulation}, we show the result of numerical simulations.
Before concluding, we extend the mechanism allowing to build the
CNOT gate to the direct generation of specific composed
gates.
\section{The system}\label{system}
\begin{figure}[ht!]
{\includegraphics[scale=0.40]{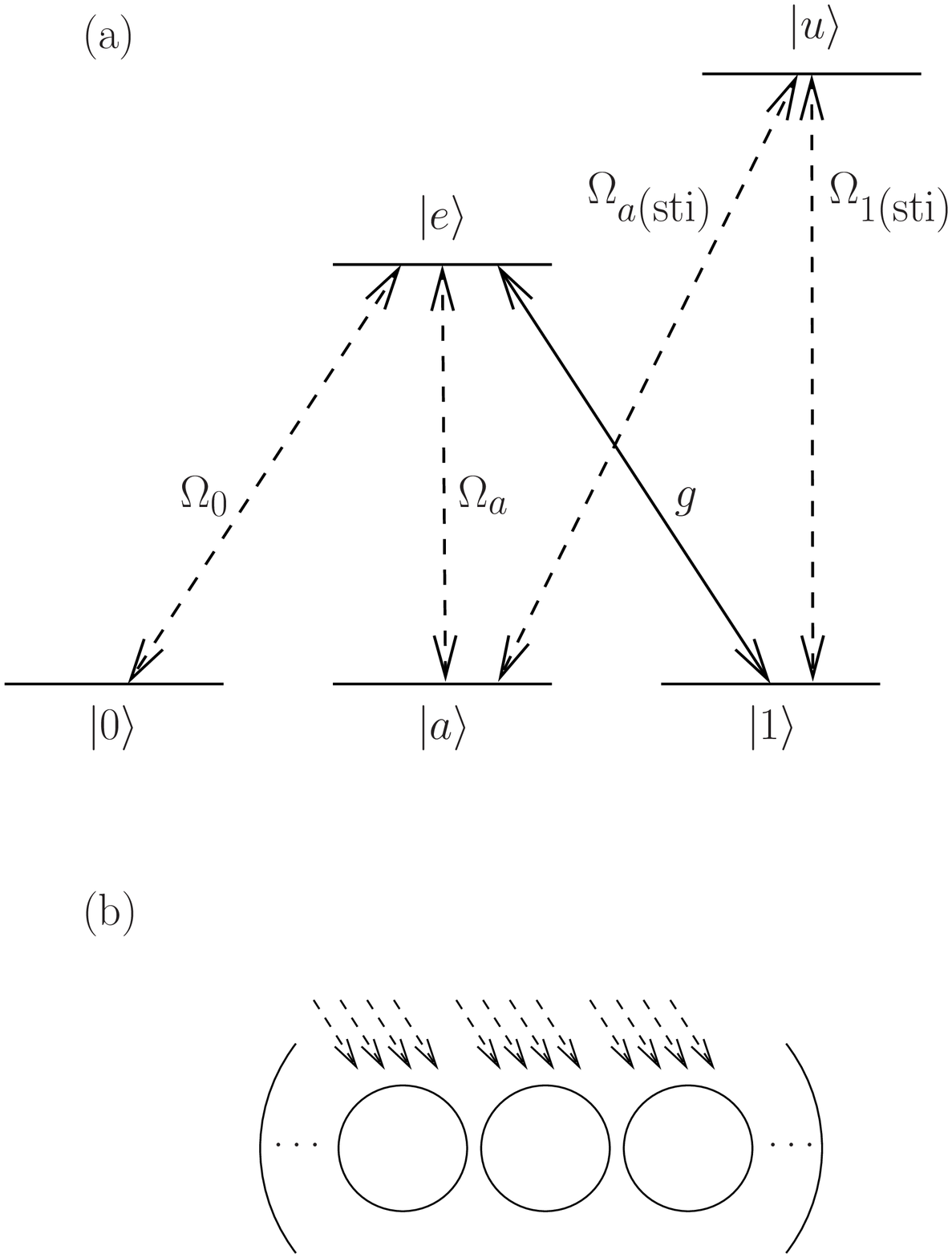}} \caption{(a)
Schematic representation of the five-level atom. The laser
(cavity) couplings are represented by dashed (full) arrows. (b)
Representation of the atomic register trapped in a single-mode
optical cavity. The atoms are represented by circles, laser fields by
arrows.} \label{scheme_atom}
\end{figure}
Although the mechanism could be realised in a system composed of
non-degenerate ground states, we use the five-level atomic system
as presented in Fig.~\ref{scheme_atom}(a), in which other gates
have also been implemented~\cite{Goto}. The three ground states
$|0\rangle,$ $|a\rangle$ and $|1\rangle$ are coupled to the
excited state $|e\rangle$ respectively by two lasers (associated
to the Rabi frequencies $\Omega_0$ and $\Omega_a$), and by a
single mode cavity (associated to the Rabi frequency $g$).
Furthermore, $|a\rangle$ and $|1\rangle$ are coupled by two
additional lasers (with
 Rabi frequencies $\Omega_{a\hbox{\scriptsize{(sti)}}}$ and
$\Omega_{1\hbox{\scriptsize{(sti)}}})$ to the upper state
$|u\rangle.$ The polarizations and the frequencies are such that
each field drives a unique transition. The atomic states
$|0\rangle$ and $|1\rangle$ represent the computational states of
the qubit. We consider that the atomic register is fixed in the
single-mode optical cavity as represented in
Fig.~\ref{scheme_atom}(b). Each atom (labeled by $k$) of the
register is driven by a set of four pulsed laser fields
$\Omega_0^{(k)}(t),$ $\Omega_a^{(k)}(t),$
$\Omega_{a\hbox{\scriptsize{(sti)}}}^{(k)}$ and
$\Omega_{1\hbox{\scriptsize{(sti)}}}^{(k)}$ and by the cavity mode
$g^{(k)}$ which is time independent.
\section{The mechanism}\label{expo}
\subsection{General strategy}
\begin{figure}[ht!]
\includegraphics[scale=0.352]{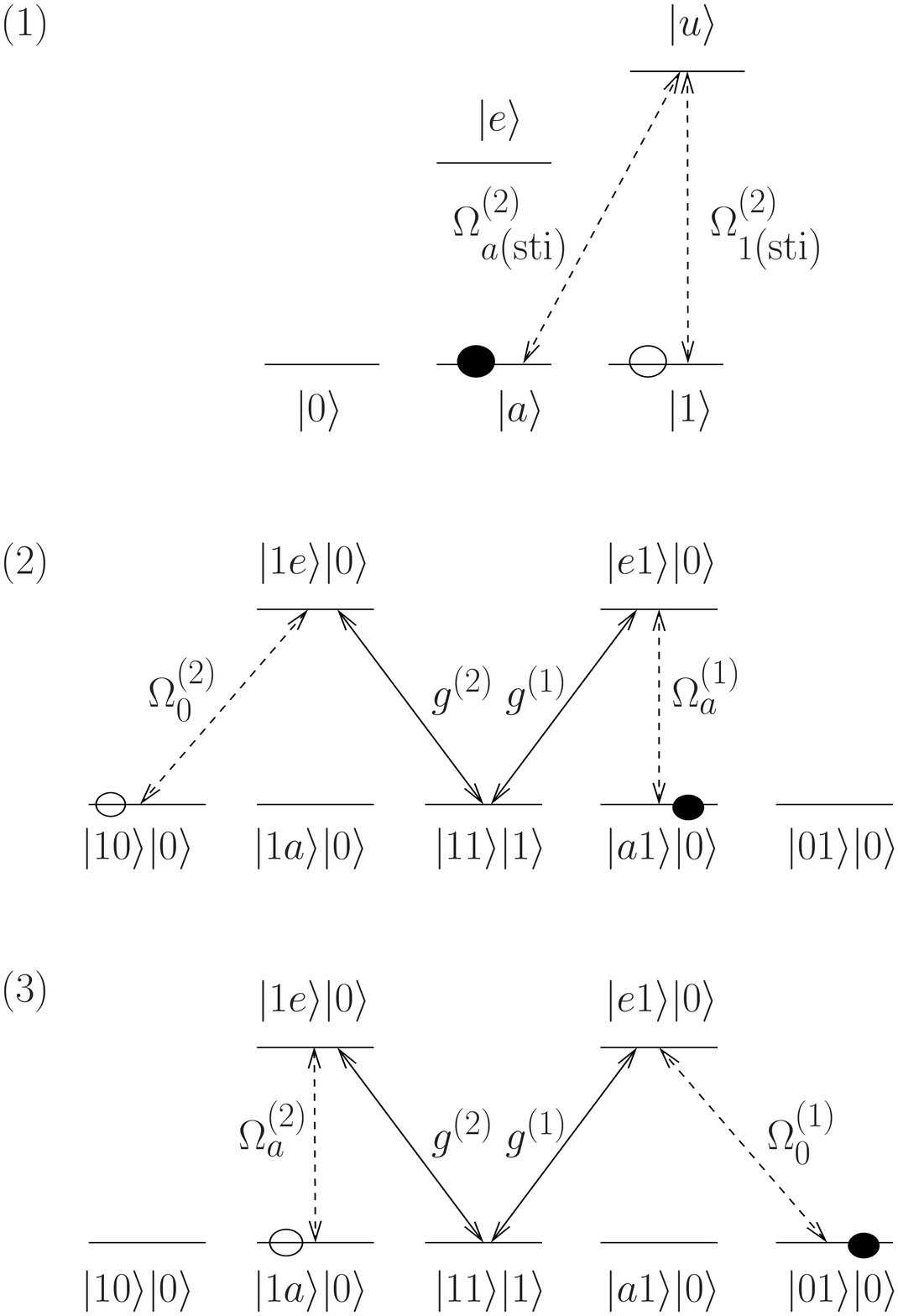}
\includegraphics[scale=0.352]{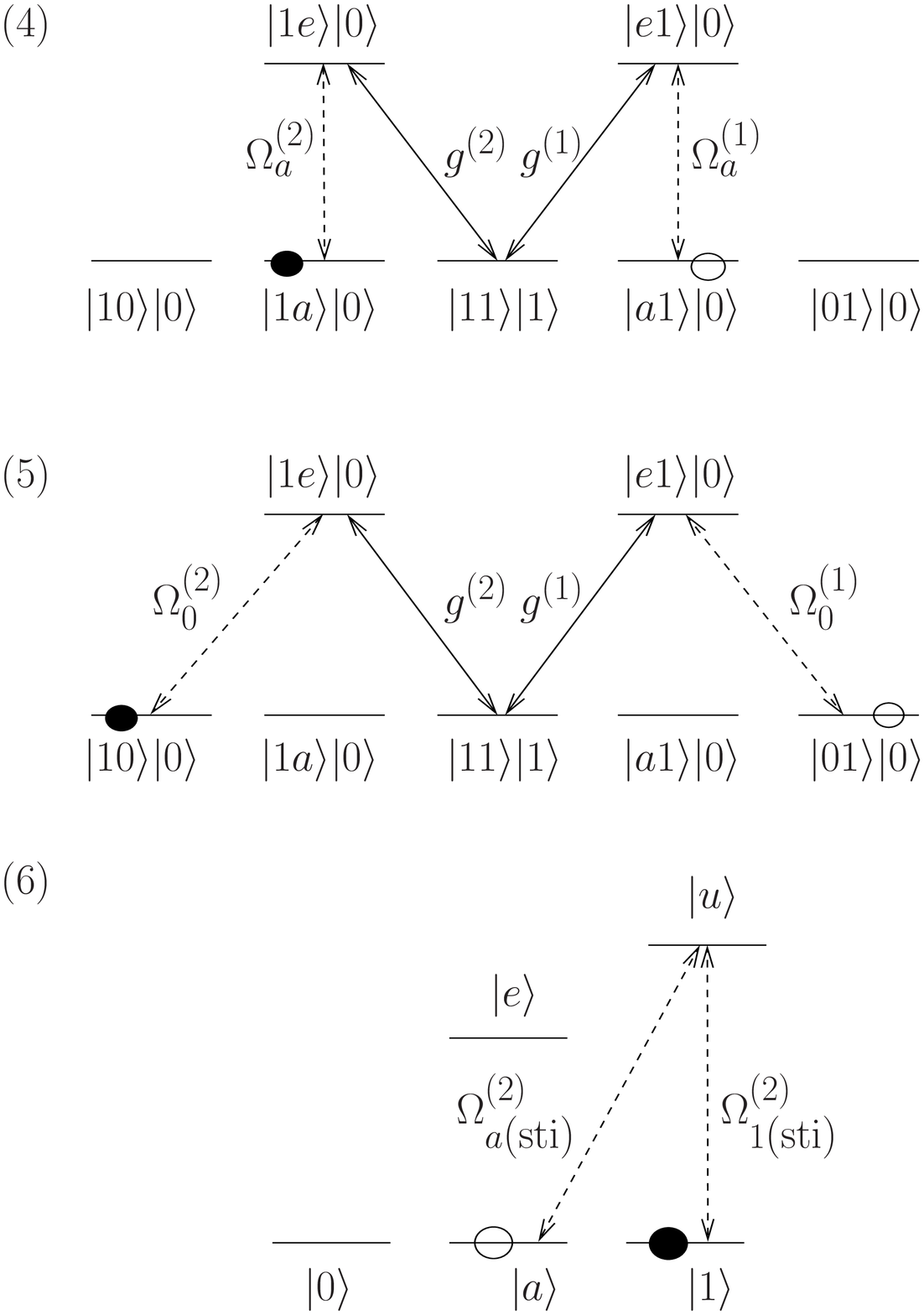}
\caption{Schematic representation of the six steps of the
cons\-truc\-tion of the CNOT gate. For each step, the initial
state is represented by an empty circle whereas the final state is
symbolized by a full black circle.} \label{fig_steps}
\end{figure}

We first recall how the CNOT gate acts.
Before the interaction
with the lasers, the initial state $|\psi_i\rangle$ of the atoms in
the cavity is defined as
\begin{equation}
\label{initial_cond}
|\psi_i\rangle= \alpha |00\rangle |0\rangle + \beta
|01\rangle |0\rangle+ \gamma |10\rangle |0\rangle+\delta
|11\rangle |0\rangle,
\end{equation}
where the labels $s_1,s_2$ of the states of the form
$|s_1s_2\rangle|0\rangle$ denote respectively the states of the
first and second atom, and $|0\rangle$ is the initial vacuum state
of the cavity-mode field. $\alpha, \beta, \gamma, \delta$ are
complex coefficients. The CNOT gate exchanges the states
$|0\rangle$ and $|1\rangle$ of the second target qubit when the
first control qubit is in state $|1\rangle$ leading to the output
state
\begin{equation}
|\psi_o\rangle= \alpha |00\rangle |0\rangle + \beta |01\rangle
|0\rangle + \gamma |11\rangle |0\rangle+\delta |10\rangle
|0\rangle.
\end{equation}
We use a simple interaction scheme to represent the proposed
mechanism for the creation of a CNOT gate (see
Fig.~\ref{fig_steps}). This mechanism is composed of six steps.
Since the state $|11\rangle|0\rangle$ is a stationary state (if
there are no photons in the cavity there cannot be any transition
from $|1\rangle$ to $|e\rangle$), we first transfer the population
of the state $|1\rangle$ of the second atom into the ancillary
state $|a\rangle$ by STIRAP. The next four steps allow to swap the
populations of the states $|10\rangle|0\rangle$ and
$|1a\rangle|0\rangle.$ The last step transfers back the population
of the ancillary state $|a\rangle$ of the second atom into the
state $|1\rangle.$ The population transfers are realised by
adiabatic passage along dark states (i.e. with no components in
the atomic excited states and a negligible component in the
excited cavity states). We thus obtain a decoherence-free method
for the creation of the CNOT gate. In the next subsection, we give
details of each step.
\subsection{Description of the steps}
The six steps summarized above are obtained as follows: \\
   \underline{Step 1}: The population of state $|1\rangle$ of the
second atom is completely transferred into $|a\rangle$ by the use
of two resonant pulses $\Omega_{a\hbox{\scriptsize(sti)}}^{(2)},$
$\Omega_{1\hbox{\scriptsize(sti)}}^{(2)}$, with relative phase
$\varphi=\pi,$
switched on and off in a counterintuitive pulse
sequence (i.e. $\Omega_{a\hbox{\scriptsize(sti)}}^{(2)}$ before
$\Omega_{1\hbox{\scriptsize(sti)}}^{(2)}$.) After this STIRAP
interaction~\cite{bergmann1, bergmann2}, the initial state
(\ref{initial_cond}) becomes
\begin{equation}
\label{state_step1}
|\psi_1\rangle=\alpha |00\rangle |0\rangle + \beta |0a\rangle |0\rangle +
\gamma |10\rangle |0\rangle+\delta |1a\rangle |0\rangle.
\end{equation}
\underline{Step 2}: The population of state $|10\rangle|0\rangle$
is transferred into $|a1\rangle|0\rangle$ with the use of the
sequence $\Omega_a^{(1)}$,$\Omega_0^{(2)}.$ This adiabatic
transfer is a non-trivial coherent process described
in~\cite{Pellizari}. It uses a five-level extended STIRAP with
constant intermediate couplings. The state (\ref{state_step1})
reads
 \begin{equation}
|\psi_2\rangle=\alpha |00\rangle |0\rangle + \beta |0a\rangle |0\rangle +
\gamma |a1\rangle |0\rangle + \delta |1a\rangle |0\rangle.
\end{equation}
\underline{Step 3}: With a similar technique, the
population of
$|1a\rangle|0\rangle$ is transferred into $|01\rangle|0\rangle$ by
the use of the counterintuitive
sequence of the two pulses
$\Omega_0^{(1)}$, $\Omega_a^{(2)}$ leading to the state
   \begin{equation}
|\psi_3\rangle=\alpha |00\rangle |0\rangle + \beta |0a\rangle |0\rangle +
\gamma |a1\rangle |0\rangle + \delta |01\rangle |0\rangle.
\end{equation}
   \underline{Step 4}: With a similar technique, the population of $|a1\rangle|0\rangle$ is
transferred into
   $|1a\rangle|0\rangle$ by the use of the sequence
   $\Omega_a^{(2)}$, $\Omega_a^{(1)}$ giving
   \begin{equation}
\label{psi_4}
|\psi_4\rangle=\alpha |00\rangle |0\rangle + \beta |0a\rangle |0\rangle +
\gamma |1a\rangle |0\rangle + \delta |01\rangle |0\rangle.
   \end{equation}
   \underline{Step 5}: With a similar technique, the population of $|01\rangle|0\rangle$ is
transferred into $|10\rangle|0\rangle$ by the use of the sequence $\Omega_0^{(2)}$, $\Omega_0^{(1)}$ in
such a way that the state (\ref{psi_4}) becomes
\begin{equation}
|\psi_5\rangle=\alpha |00\rangle |0\rangle + \beta
|0a\rangle|0\rangle +
\gamma |1a\rangle |0\rangle+\delta |10\rangle
|0\rangle,
   \end{equation}
  \underline{Step 6}: The population of the state $|a\rangle$ of the
second atom is transferred back by STIRAP into
$|1\rangle$ by the use of the sequence of pulses
$\Omega_{1\hbox{\scriptsize{(sti)}}}^{(2)}$,
$\Omega_{a\hbox{\scriptsize{(sti)}}}^{(2)}$ with relative phase $\varphi=\pi.$ As a result,
the system is in state
   \begin{equation}
|\psi_6\rangle=\alpha |00\rangle |0\rangle + \beta
|01\rangle|0\rangle + \gamma |11\rangle |0\rangle+\delta
|10\rangle|0\rangle,
   \end{equation}
which coincides with the output state of the CNOT gate.
\subsection{Calculation of the instantaneous eigenstates}
We calculate the instantaneous eigenvectors connected with the
initial condition and that are thus adiabatically followed by the
dynamics when the two atoms interact with two laser fields and the
cavity-mode. We show that they are dark states with no component
in the atomic excited states and a negligible component in the
excited
cavity states.\\
We give the details of steps (1)-(6) first and next (2)-(3)-(4)-(5). \\
The steps (1) and (6) are the well known STIRAP
process~\cite{bergmann1,bergmann2}. The dynamics follows the dark
state
\begin{equation}
|\phi_{\hbox{\scriptsize{(sti)}}}\rangle
\propto
\Omega_{\hbox{\scriptsize{a(sti)}}}^{(2)}|1\rangle-e^{i\varphi}
\Omega_{\hbox{\scriptsize{1(sti)}}}^{(2)}|a\rangle
\end{equation}
(where
$\varphi$ is the relative phase of the pulses $
\Omega_{\hbox{\scriptsize{1(sti)}}}^{(2)}$ and $
\Omega_{\hbox{\scriptsize{a(sti)}}}^{(2)}$) that transfers population
from $|1\rangle$ $(|a\rangle)$ to $|a\rangle$ $(|1\rangle)$ with a
counterintuitive pulse sequence. We choose the phase $\varphi=\pi$ to avoid
a minus sign on the states $|a\rangle$ and $|1\rangle$ of the second
atom after the steps (1) and (6) respectively.\\
Concerning the intermediate steps, since the lasers do not couple the atomic
state $|1\rangle$, the state
$|\phi_1^{(1)}\rangle=|11\rangle|0\rangle$ (defining one dimensional Hilbert space ${\cal H}_1$) of the
initial
condition (\ref{initial_cond}) is decoupled from the other ones. The other
states of (\ref{initial_cond}) are connected to two orthogonal decoupled
subspaces denoted ${\cal H}_{7}$ and ${\cal H}_{16}$, respectively
spanned by the states
\begin{eqnarray}
{\cal H}_7&=&\{ |01\rangle|0\rangle,|10\rangle|0\rangle,
|1a\rangle|0\rangle, |a0\rangle|0\rangle,\nonumber\\
&& |1e\rangle|0\rangle, |e1\rangle|0\rangle,
|11\rangle|1\rangle\}, \label{B7}
\end{eqnarray}
and \begin{eqnarray} \nonumber && {\cal
H}_{16}=\{|00\rangle|0\rangle, |0a\rangle|0\rangle,
|01\rangle|1\rangle,
|0e\rangle|0\rangle, |a0\rangle|0\rangle,  \\
&&\nonumber  |aa\rangle|0\rangle,|a1\rangle|1\rangle,
|ae\rangle|0\rangle, |1e\rangle|1\rangle, |e0\rangle|0\rangle,
|10\rangle|1\rangle, \\
&& |1a\rangle|1\rangle,|11\rangle|2\rangle, |ea\rangle|0\rangle,
|e1\rangle|1\rangle,|ee\rangle|0\rangle\}. \label{B16}
\end{eqnarray}
For each step, one ground state $|0\rangle$ or $|a\rangle$ of each
atom is coupled by
a laser field to the excited state, while the other one is not
coupled to the excited state. To describe the calculation of the
instantaneous eigenstates for the four steps, we introduce the
following notation : the state coupled by a laser field is labeled
$|L^{(i)}\rangle$ ($|0^{(i)}\rangle$ or
$|a^{(i)}\rangle$) and the non-coupled state
$|N^{(i)}\rangle$ ($|a^{(i)}\rangle$ or
$|0^{(i)}\rangle$). The index $i=1,2$ labels the atom $i$. In the full
Hilbert space $\mathcal{H}=\mathcal{H}^{(1)} \otimes
\mathcal{H}^{(2)} \otimes \mathcal{F}$ with $\mathcal{H}^{(i)}$
the Hilbert space associated to the atom $i$ and $\mathcal{F}$ the
Fock space, the Hamiltonian (in units such that $\hbar=1$) reads
in the rotating wave approximation
\begin{eqnarray}
\label{Ham_gen}
&&H(t) =\omega_c a^\dagger a+ \omega_e
|e^{(1)}\rangle\langle e^{(1)}| +\omega_e
|e^{(2)}\rangle\langle e^{(2)}|\nonumber\\
\nonumber &&   +\left(\Omega^{(1)}(t) e^{-i\omega t}
|e^{(1)}\rangle\langle L^{(1)}| +g^{(1)}a
|e^{(1)}\rangle\langle 1^{(1)}|
+h.c.\right) \\
&& +\left(\Omega^{(2)}(t) e^{-i\omega t} |e^{(2)}\rangle\langle
L^{(2)}|+g^{(2)}a |e^{(2)}\rangle\langle
1^{(2)}|+h.c.\right)\nonumber\\
\end{eqnarray}
where $a$ $(a^\dagger)$ is the annihilation (creation) operator
for the cavity mode, $\omega$ ($\omega_c$) is the frequency of the
laser field (cavity mode) and $\omega_e$ is the energy of the
excited state (the energy reference is taken for the ground
states: $\omega_0=\omega_a=\omega_1=0$). We consider resonant fields: $\omega_e=\omega=\omega_c$. $\Omega^{(i)}(t)$ and
$g^{(i)}$ are the Rabi frequencies associated to the laser pulse
and to the cavity respectively for the atom $i$.
($\Omega^{(i)}(t)$ corresponds to $\Omega_0^{(i)}(t)$ or
$\Omega_a^{(i)}(t)$, depending on the ground state $|0\rangle$ or
$|a\rangle$ of the atom $i$ coupled by the laser.) The dynamics is
determined by the Schr\"odinger equation
$i\frac{\partial}{\partial t} \psi(t)=H(t)\psi(t).$ The
Hamiltonian in the interaction picture
\begin{subequations}
\begin{equation}\label{Heff}
H_{\hbox{\scriptsize{I}}}(t)=T^{\dagger}(t)H(t) T(t) -i
T^\dagger(t) \frac{dT}{dt}(t)
\end{equation}
with
\begin{equation}\label{def_TI}
   T(t)= e^{-i\omega t \left( a^{\dagger}a+|e^{(1)}\rangle\langle e^{(1)}|+
   |e^{(2)}\rangle\langle e^{(2)}|\right)}
\end{equation}
\end{subequations}
reads
\begin{eqnarray}
&H_{\hbox{\scriptsize{I}}}(t)&=\Omega^{(1)}(t)|e^{(1)}\rangle\langle
L^{(1)}|
+g^{(1)}a |e^{(1)}\rangle\langle
1^{(1)}|\qquad\nonumber\\
&& +\Omega^{(2)}(t)|e^{(2)}\rangle\langle
L^{(2)}|+g^{(2)}a |e^{(2)}\rangle\langle
1^{(2)}|\nonumber\\
&& +h.c.
\end{eqnarray}
The dynamics is therefore determined by
\begin{equation}
\label{schro_heff} i \frac{\partial}{\partial t}
\phi(t)=H_{\hbox{\scriptsize{I}}}(t)\phi(t)
\end{equation}
with
\begin{equation}
\label{connection} \psi(t)=T(t)\phi(t).
\end{equation}
We remark that the transformation $T$ does not change the initial state
(\ref{initial_cond}). The Hamiltonian $H_{\hbox{\scriptsize{I}}}(t)$
is block-diagonal;
three blocks connected to the initial state (\ref{initial_cond}) have to
be considered:
\begin{equation}
H_{\hbox{\scriptsize{I}}}(t)=\left[
\begin{array}{ccc}
H_{1}=0 & 0 & 0 \\
0 & H_{7}(t) & 0 \\
0 & 0 & H_{16}(t)%
\end{array}%
\right]
\end{equation}
with $H_d(t)$ acting in the $d$-dimensional subspace ${\cal
H}_{d}$ generated by the set of states defined in Eqs. (\ref{B7})
and (\ref{B16}). The adiabatic evolution of the initial state
(\ref{initial_cond}) is completely described by the dark states of
the Hamiltonian $H_{\hbox{\scriptsize{I}}}$, labeled
$\phi_{d}^{(k)}(t)\in {\cal H}_{d}$ (with k the index of
degeneracy of ${\cal H}_{d}$). These dark states are instantaneous
eigenstates that don't have any components on the atomic excited
states. They are associated to null eigenvalues. Although these
dark states are degenerate, they evolve without any geometric
phase. One can easily check that all the elements contributing to
this geometric phase \cite{Zee},
$\langle\phi_d^{(k^\prime)}(s)|\frac{d}{ds}|\phi_d^{(k)}(s)\rangle$,
are null during the dynamics since for $k=k^\prime,$ the phase of
the lasers is constant for each step (as in standard STIRAP) and
for $k\neq k^\prime,$ the dark states belong to orthogonal
subspaces. Therefore, according to the adiabatic theorem, the
dynamics follows the dark states initially connected to each
component of the initial state (\ref{initial_cond})
\begin{equation}
|\psi(t)\rangle \approx
T(t)\sum_{d, k}
c_d^k|\phi_{d}^{(k)}(t)\rangle,
\end{equation}
with the coefficients
\begin{eqnarray}
\nonumber
c_d^k & = &\langle\phi_d^{(k)}(t_i)|T^{\dagger}(t_i)|\psi(t_i)\rangle \\
& = & \langle\phi_d^{(k)}(t_i)|\psi(t_i)\rangle.
\end{eqnarray}
We have thus to determine the
instantaneous eigenstates. In the
subspace ${\cal H}_7$, the states
$|\phi_7^{(1)}\rangle=|N^{(1)}1\rangle |0\rangle$ and
$|\phi_7^{(2)}\rangle=|1N^{(2)}\rangle
|0\rangle$ are not
coupled to the initial state (\ref{initial_cond}) and do not participate
in the dynamics. Only the atomic dark state~\cite{Pellizari}:
\begin{eqnarray}
\label{dark3}  |\phi_{7}^{(3)}\rangle&\propto&g^{(1)}\Omega^{(2)}
|L^{(1)}1\rangle|0\rangle+
g^{(2)}\Omega^{(1)}|1L^{(2)}\rangle
|0\rangle
\nonumber\\
&& -\Omega^{(1)}\Omega^{(2)}|11\rangle |1\rangle
\end{eqnarray}
(where the normalisation coefficient has been omitted)
participates to the dynamics. The second step, associated to
$L^{(1)}\equiv a$,
$L^{(2)}\equiv 0$,
$\Omega^{(1)}\equiv\Omega_a^{(1)}$,
$\Omega^{(2)}\equiv\Omega_0^{(2)}$ leads to the initial and final
connections symbolically written as
$|10\rangle|0\rangle\rightarrow|\phi_{7}^{(3)}\rangle\rightarrow|a1\rangle|0\rangle$
(see Fig.~3). The third, fourth and fifth steps give
respectively the connections
$|1a\rangle|0\rangle\rightarrow|\phi_{7}^{(3)}\rangle\rightarrow|01\rangle|0\rangle$,
$|a1\rangle|0\rangle\rightarrow|\phi_{7}^{(3)}\rangle\rightarrow|1a\rangle|0\rangle$,
and
$|01\rangle|0\rangle\rightarrow|\phi_{7}^{(3)}\rangle\rightarrow|10\rangle|0\rangle$.
We determine four atomic dark states in the subspace ${\cal H}_{16}$
connected to the component $|00\rangle|0\rangle$ of the
initial condition (\ref{initial_cond}):
\begin{subequations}
\begin{eqnarray}\label{dark_1}
|\phi_{16}^{(2)}\rangle&\propto&\Omega^{(2)}
|N^{(1)}1\rangle|1\rangle-g^{(2)}|N^{(1)}
L^{(2)}\rangle
|0\rangle\\
  |\phi_{16}^{(3)}\rangle&\propto& \Omega^{(1)}
|1N^{(2)}\rangle|1\rangle-
g^{(1)}|L^{(1)}N^{(2)}\rangle
|0\rangle,\\
|\phi_{16}^{(4)}\rangle&=&|N^{(1)}N^{(2)}\rangle
|0\rangle,\\
|\phi_{16}^{(5)}\rangle&\propto& g^{(1)}g^{(2)}\sqrt{2}
|L^{(1)}L^{(2)}\rangle|0\rangle-
g^{(2)}\Omega^{(1)}\sqrt{2}|1L^{(2)}\rangle
|1\rangle \nonumber\\
&&-g^{(1)}\Omega^{(2)}\sqrt{2}|L^{(1)}1\rangle
|1\rangle+ \Omega^{(1)}\Omega^{(2)}|11\rangle |2\rangle.
\end{eqnarray}
\end{subequations}
We remark that the state $|00\rangle|0\rangle$ is connected initially
and finally
to the dark state $|\phi_{16}^{(n)}\rangle$ at the $n^{\text{th}}$
step.
Since the dynamics follows atomic dark
states, the excited atomic state is never populated (in the
adiabatic limit). Moreover, the projections of the dark states on
the excited cavity photon states can be made negligible if $g^{(i)} \gg
\Omega^{(i)}$~\cite{malinovky}. In this case, the mechanism we
propose is a decoherence-free method in the sense that the process
is not sensitive to spontaneous emission from the atomic excited
states nor to the lifetime of photons in the optical cavity.
\section{Numerical validation}\label{simulation}

\begin{figure}[ht!]
{\includegraphics[scale=0.55]{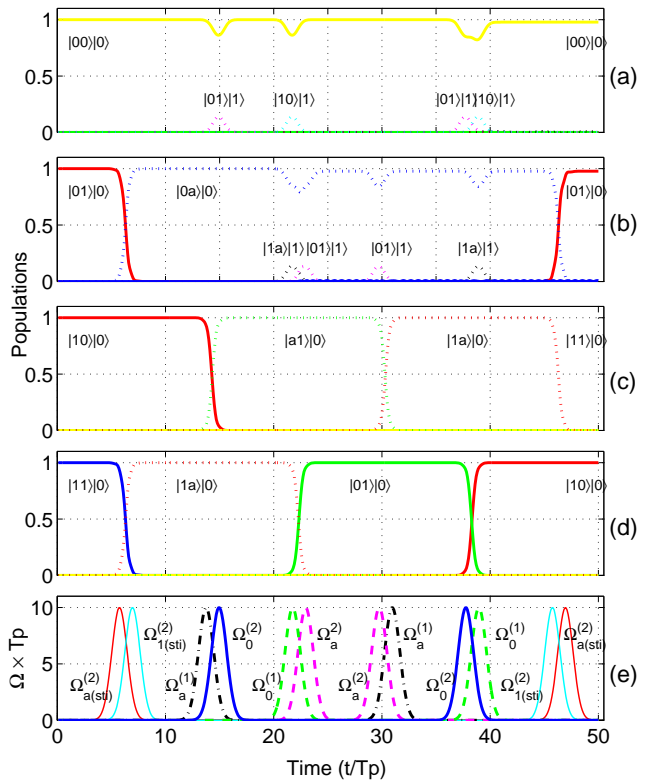}}
\caption{(Colour online) Temporal evolution of initial conditions (a)
$|00\rangle|0\rangle,$ (b) $|01\rangle|0\rangle$ (c) $|10\rangle|0\rangle$ and
(d) $|11\rangle|0\rangle.$ The states which are populated during the
interaction with the pulses or between two steps are indicated.
(e) Temporal profile of the Rabi frequencies. The parameters used
are $\Omega_{\max}T_p=10,$ $g T_p=25.$ The delay between two pulses in a step is $1.2T_p.$}
\label{time_evolution}
\end{figure}

We present the numerical validation of the mechanism proposed for
the construction of the CNOT gate.\\
We show in Fig.~\ref{time_evolution} the time evolution of four
initial states: in (a) and (b) the population of initial states
$|00\rangle|0\rangle$ and $|01\rangle|0\rangle$ respectively stays
in these states after the interaction with the twelve pulses since
the control qubit is in state $|0\rangle$, in (c) and (d) the
population of initial states $|01\rangle|0\rangle$ and
$|11\rangle|0\rangle$ are exchanged. In (e), we show the Rabi
frequencies associated to the pulses. The laser Rabi frequencies
are all chosen of the form $\Omega(t)=\Omega_{\max}
e^{-\left(\frac{t}{T_p}\right)^2}.$ The steps (1) and (6) of the
mechanism can be explained by the standard STIRAP
technique~\cite{bergmann1,bergmann2}. The other steps involve the
two atoms and the cavity using an adiabatic transfer which is a
five-level extended STIRAP with constant intermediate
couplings~\cite{Pellizari}. The couplings have to satisfy
$\Omega_{\max} T_p, gT_p\gg 1$ to fulfill the adiabatic
conditions. The delay between two pulses of the same step is
chosen equal to $1.2\,T_p$ to minimize the non-adiabatic
losses~\cite{vitanov}. Moreover, the condition $g \gg
\Omega_{\max}$ has to be satisfied such that the cavity mode is
negligibly populated during the interaction with the pulses.

\section{Discussion}
In the optical domain, one can give an estimate of the
relevant parameters. Taking into account the losses of the cavity
(characterized by the decay rate $\kappa$ of the cavity field) and
of the excited states (of lifetime $\tau$), we have to satisfy the
adiabatic conditions: $\Omega_{\max} T_p,gT_p\gg 1$ and
$(\Omega_{\max}T_p)^2, (gT_p)^2 \gg \kappa T_p, T_p/\tau$. The
latter is satisfied for $g, \Omega_{\max} \gg \kappa, 1/\tau.$ For
a typical pulse duration of $T_p=50$ns, we use
$\Omega_{\max}=1.2\times$10$^8$s$^{-1}$, which is achievable
experimentally (see for instance Ref. \cite{vitanov_01}). We use a
cavity coupling $g=5.2\times$10$^8$s$^{-1}$, more than four times
larger than $\Omega_{\max}$ to have a small population in the
cavity field. Such a strong coupling has been recently achieved in
experiments with atoms in an optical cavity trapped with a
duration of the order of one second \cite{keever_03, miller_05}.
For a realistic decay rate $(\kappa=1\times10^7$s$^{-1})$ of the
cavity, the numerical simulation of the proposed process gives:
(i) 80\% of the population of states $|00\rangle|0\rangle$ and
$|01\rangle|0\rangle$ are left on these states and (ii) the
exchange of the population between $|10\rangle|0\rangle$ and
$|11\rangle|0\rangle$ is of the order of 90\%. For a decay rate
$(\kappa=1\times10^6$s$^{-1})$, we would obtain that 96\% of the
population of the states $|00\rangle|0\rangle$ and
$|01\rangle|0\rangle$ are preserved, and 92 \% of the population
of $|10\rangle|0\rangle$ and $|11\rangle|0\rangle$ are exchanged.
This analysis shows that the mechanism could be implemented with
an observable efficiency with the currently available technology.
Longer cavity photon lifetimes and/or larger cavity couplings
would give a very good efficiency.

We remark that the pulse $\Omega_a^{(2)}$ is used two times
successively in the steps (3) and (4). These two pulses can thus
be replaced by a single pulse. The
process then requires the use of only eleven pulses.\\
By manipulating the phase of the pulses, the technique proposed in
this paper can be extend to the direct preparation of the
composition of elementary gates. Indeed, if instead of taking a
phase difference equal to $\pi$ between the first and the
second laser for the steps (1) and (6) and zero otherwise, we add
an arbitrary relative phase $\varphi_{(n)}$ in step $(n)$, the
proposed mechanism leads to the following gate composition:
\begin{eqnarray}
\label{composition_phases_CNOT}
\nonumber
& & Ph^{(2)}(\varphi_{(6)}) \circ
\hbox{C-}phase{(\varphi_{(2)}+\varphi_{(4)})} \circ \hbox{CNOT}\circ
\\
& & \hbox{C-}phase{(\varphi_{(3)}+\varphi_{(5)})} \circ
Ph^{(2)}(\varphi_{(1)}).
\end{eqnarray}
Similarly, the technique we proposed to build the SWAP gate
in Ref. \cite{sangouard} leads to the composition
\begin{equation}
Ph^{(1)}(\alpha)\circ \hbox{SWAP}\circ Ph^{(1)}(\beta)\circ \hbox{C-}phase(-\alpha-\beta)
\end{equation}
where $\alpha,$ $\beta$ are functions of the static relative
phases of the laser pulses. 
\section{Conclusion}
In this paper, we have proposed a mechanism for the construction
of a CNOT gate. This technique requires the use of one cavity and
eleven pulses. It is robust against variations of amplitude and
duration of the pulses and of the delay between the pulses.
Moreover, it constitutes a decohence-free method in the sense that
the excited atomic states with short life times are not populated
in the adiabatic limit, and the cavity mode is negligibly
populated during the process. This technique can be also
an alternative to the composition of many elementary gates by a
direct construction of specific gates, which could have potential
applications for the fast realisation of some algorithms.
\begin{acknowledgments}
N.S. acknowledges support from the EU network QUACS under Contract No. HPRN-CT-2002-0039 and from La Fondation Carnot.
\end{acknowledgments}


\begin{references}
\bibitem{explanation} The
set $\{U, CNOT\}$ is a universal set and the $CNOT$ gate can be
obtained by the composition of two Hadamard gates and one
C-$phase(\pi)$ gate.
\end{references}
\end{document}